\documentclass[conference]{IEEEtran}

\usepackage[utf8]{inputenc}
\usepackage[T1]{fontenc}
\usepackage{amsmath,amssymb,amsthm,mathtools}
\usepackage{stmaryrd}  
\usepackage{bbm}
\usepackage{braket}
\usepackage{booktabs}

\usepackage{tikz}
\usetikzlibrary{arrows.meta,positioning,decorations.pathmorphing,decorations.pathreplacing,calc,backgrounds,matrix,shapes.geometric,shapes.misc}
\usepackage{cite}

\newtheorem{theorem}{Theorem}
\newtheorem{lemma}{Lemma}
\newtheorem{proposition}{Proposition}
\newtheorem{corollary}{Corollary}
\newtheorem{definition}{Definition}

\newtheorem{remark}{Remark}

\newcommand{\ZZ}{\mathbb{Z}}
\newcommand{\Prob}{\mathbb{P}}

\newcommand{\1}{\mathbbm{1}}
\newcommand{\DSR}{\mathrm{DSR}}
\newcommand{\CSR}{\mathrm{CSR}}
\newcommand{\HCSR}{\mathrm{HCSR}}
\newcommand{\F}{\mathbb{F}}

\DeclareMathOperator{\supp}{supp}

\begin{document}

\title{Strip-Symmetric Quantum Codes for Biased Noise:\\ 
Z-Decoupling in Stabilizer and Floquet Codes}

\author{
\IEEEauthorblockN{Mohammad Rowshan}
\IEEEauthorblockA{
University of New South Wales (UNSW)\\Sydney, Australia\\
Email: mrowshan@ieee.org}
}

\maketitle

\begin{abstract}
Bias-tailored codes such as the XZZX surface code and the domain wall color code achieve high dephasing-biased thresholds because, in the infinite-bias limit, their Z syndromes decouple into one-dimensional repetition-like chains; the X\(^3\)Z\(^3\) Floquet code shows an analogous strip-wise structure for detector events in spacetime. We capture this common mechanism by defining \emph{strip-symmetric biased codes}, a class of static stabilizer and dynamical (Floquet) codes for which, under pure dephasing and perfect measurements, each elementary Z fault is confined to a strip and the Z-detector–fault incidence matrix is block diagonal. For such codes the Z-detector hypergraph decomposes into independent strip components and maximum-likelihood Z decoding factorises across strips, yielding complexity savings for matching-based decoders. We characterise strip symmetry via per-strip stabilizer products (a \(\mathbb{Z}_2\) 1-form symmetry), place XZZX, the domain wall color code, and X\(^3\)Z\(^3\) in this framework, and introduce synthetic strip-symmetric detector models and domain-wise Clifford constructions that serve as design tools for new bias-tailored Floquet codes.
\end{abstract}


\section{Introduction}
Quantum error correction (QEC) is essential for scalable quantum computation, but among topological codes the surface code is suboptimal under dephasing-biased noise. Bias-tailored codes have achieved major improvements, notably the XZZX surface code~\cite{tuckett2018ultrahigh,tuckett2019tailoring,bonilla2021xzzx} and the domain wall color code~\cite{tiurev2024domainwall}. In the infinite-bias limit, Z errors in these static codes align along preferred directions and their Z syndromes decouple into one-dimensional repetition chains with line symmetries, enabling efficient biased decoding and high thresholds~\cite{bonilla2021xzzx,bonilla2021xzzxsupp,tiurev2024domainwall,domainwall_supplement}. In parallel, dynamical (Floquet) codes are defined by periodic sequences of local Pauli measurements rather than a fixed stabilizer group~\cite{hastings2021dynamical,fu2025dynamical}. Within this setting, the X$^3$Z$^3$ Floquet code~\cite{setiawan2025x3z3,setiawan2025x3z3colab}, a Clifford deformation of a CSS Floquet code on the honeycomb lattice, achieves strong dephasing-biased performance~\cite{setiawan2025x3z3} and exhibits a stabilizer-product symmetry: in the infinite-bias limit, certain plaquette products enforce a conservation law along vertical domains, so Z-type defects on each domain appear in pairs and can be decoded domain-wise by MWPM~\cite{setiawan2025x3z3,setiawan2025x3z3supp}. These examples suggest a common structural mechanism—strip-wise decoupling of the Z-decoding problem under strong Z bias—that has not yet been formalised as a general property of biased static and dynamical codes.
We formalise this mechanism via \emph{strip-symmetric biased codes}, a class of static stabilizer and dynamical (Floquet) codes with a partition of qubits and detectors into strips such that, under pure dephasing and perfect measurements, each elementary Z fault and a per-strip parity constraint are confined to a single strip. For this class the Z-detector incidence matrix can be permuted into block-diagonal form, so the Z-detector hypergraph splits into independent strip components and maximum-likelihood Z decoding factorises across strips, yielding complexity savings for matching-based decoders. We characterise strip symmetry by per-strip stabilizer products, i.e.\ a $\mathbb{Z}_2$ 1-form symmetry, which unifies the line symmetries in XZZX and domain wall color codes~\cite{bonilla2021xzzx,tiurev2024domainwall,domainwall_supplement} with the domain-wise conservation laws in X$^3$Z$^3$~\cite{setiawan2025x3z3,setiawan2025x3z3supp}, and we show that domain-wise Clifford deformations of CSS Floquet codes automatically yield strip-symmetric Floquet codes. As benchmarks we introduce synthetic strip-symmetric detector models—diagonal-, column-, and half-density column-strip repetition ($\DSR$, $\CSR$, $\HCSR$)—whose Z-detector graphs are stacks of repetition chains and serve as design testbeds; in contrast to the code-specific analyses of~\cite{bonilla2021xzzx,bonilla2021xzzxsupp,tiurev2024domainwall,domainwall_supplement,setiawan2025x3z3,setiawan2025x3z3supp}, our framework gives a unified structural notion with general ML-factorisation and complexity guarantees.

Due to space limits, proofs are omitted but provided in \cite{rowshan2026strip}. 
\section{Preliminaries}
\label{sec:prelim}

\subsection{Static stabilizer and dynamical codes}

We use standard stabilizer codes and the dynamical-code framework of
Fu and Gottesman~\cite{fu2025dynamical}. 
A stabilizer code $Q(S)$ with parameters $\llbracket n,k,d\rrbracket$ is
specified by a set $S=\{s_1,\dots,s_{n-k}\}$ of commuting $n$-qubit Pauli
operators generating a stabilizer group $\langle S\rangle\subseteq\mathcal{P}_n$.
The codespace is
\[
  \mathcal{C} := \{ \ket{\psi} : s\ket{\psi}=\ket{\psi}\ \forall\,s\in S\}.
\]
Logical Pauli operators are elements of the normaliser $N(S)$ not in $\langle
S\rangle$, and the distance is
\[
  d := \min\{\operatorname{wt}(P) : P\in N(S)\setminus\langle S\rangle\}.
\]

A \emph{dynamical code} is given by a sequence $(M_i)_{i\in\mathbb{Z}}$ of
commuting measurement sets $M_i\subseteq\mathcal{P}_n$. The instantaneous
stabilizer group $S_i$ after round $i$ is obtained from $S_{i-1}$ and the
outcomes of the measurements in $M_i$ via stabilizer update rules, and defines
a codespace $\mathcal{C}_i$. A \emph{Floquet} code is a dynamical code with a
finite period $T$ such that $M_{i+T}=M_i$ for all $i$.
A static stabilizer code corresponds to the trivial periodic sequence
$M_i=S$ for all $i$ with $T=1$, so all definitions below apply uniformly to static and
Floquet codes.

\subsection{Biased Pauli noise}

We assume an i.i.d.\ single-qubit biased Pauli channel
\[
  \mathcal{N}(\rho)
  = (1-p)\rho
    + p_X X\rho X
    + p_Y Y\rho Y
    + p_Z Z\rho Z,
\]
with $p_X,p_Y,p_Z\ge 0$ and $p_X+p_Y+p_Z=p$. The Z-bias is
\[
  \eta := \frac{p_Z}{p_X+p_Y},
\]
so $\eta\gg 1$ corresponds to strongly dephasing-biased noise. In the \emph{pure dephasing} limit, we set $p_X=p_Y=0$ and $p_Z=p$.
The infinite-bias regime $\eta\to\infty$ is standard in the analysis of bias-tailored codes such as the XZZX surface code, the domain wall color code, and the X$^3$Z$^3$ Floquet code. 

\subsection{Detectors and Z-detector (hyper)graphs}

We work in the detector-cell picture for static and dynamical
codes. 
Let $(M_i)$ be the
measurement sequence.

A \emph{detector} is a Pauli observable
\[
  D = \prod_{(P,i)\in\mathcal{I}} m(P,i),
\]
where $\mathcal{I}$ is a finite set of measurement events $(P,i)$ with
$P\in M_i$ and outcomes $m(P,i)\in\{\pm 1\}$. We choose detectors so that, in
the absence of noise, $D$ has a fixed outcome (conventionally $+1$); an
outcome $-1$ indicates an excitation associated with $D$. Let $\mathcal{D}$ be
a chosen set of independent detectors whose outcomes form the (Z) syndrome.

Fix a noise model and a set of elementary Z faults $\mathcal{F}$ (e.g.\
single-qubit Z or Z-type measurement errors); each fault $f\in\mathcal{F}$
flips a subset $\partial(f)\subseteq\mathcal{D}$ of detectors.

\begin{definition}[Z-detector hypergraph and incidence matrix]
The \emph{Z-detector hypergraph} is
\[
  (\mathcal{D},\mathcal{E}_Z),\qquad
  \mathcal{E}_Z := \{\partial(f): f\in\mathcal{F}\},
\]
with vertices given by detectors and one hyperedge per fault.
Indexing detectors by $i$ and faults by $\ell$, the Z-detector incidence
matrix $H_Z$ is the binary matrix with entries
\[
  (H_Z)_{i\ell} =
  \begin{cases}
    1, &\text{if detector $i$ is flipped by fault $\ell$},\\
    0, &\text{otherwise}.
  \end{cases}
\]
Writing $e\in\{0,1\}^{|\mathcal{F}|}$ for the fault vector and
$s\in\{0,1\}^{|\mathcal{D}|}$ for detector outcomes (mod $2$), we have
\[
  H_Z e = s \pmod 2.
\]
\end{definition}

\begin{figure}[t]
  \centering
  \begin{tikzpicture}[scale=0.9,
    detector/.style={circle,draw,inner sep=1pt,minimum size=5mm,fill=white},
    edge2/.style={very thick,blue!70},
    hyperedge/.style={rounded corners=8pt,very thick,green!70,fill=green!10},
    lab/.style={font=\small}
  ]

    \node[detector] (d1) at (0,1.6) {$d_1$};
    \node[detector] (d2) at (1.6,2.4) {$d_2$};
    \node[detector] (d3) at (3.2,1.6) {$d_3$};
    \node[detector] (d4) at (1.0,0.3) {$d_4$};
    \node[detector] (d5) at (2.6,0.3) {$d_5$};

    \draw[edge2] (d1) -- (d2);
    \draw[edge2] (d2) -- (d3);

    \node[lab,blue!70] at (0.6,2.6) {$f_1$};
    \node[lab,blue!70] at (2.6,2.6) {$f_2$};

    \begin{scope}[on background layer]
      \path[hyperedge]
        ($(d3)+(-0.4,0.4)$) --
        ($(d3)+(0.4,0.4)$) --
        ($(d5)+(0.5,0)$) --
        ($(d4)+(-0.5,0)$) -- cycle;
    \end{scope}
    \node[lab,green!70,anchor=west] at (3.4,1.0) {$f_3$};

    \node[lab,align=center] at (1.6,-0.7)
      {Z-detector hypergraph\\vertices: detectors $d_i$,\\hyperedges: faults $f_\ell$};

    \begin{scope}[shift={(6.5,0.7)}]
      \node[lab] at (0,2.4) {$H_Z$};

      \matrix[matrix of math nodes,
              nodes in empty cells,
              left delimiter={[},
              right delimiter={]}] (M) {
        1 & 0 & 0 \\  
        1 & 1 & 0 \\  
        0 & 1 & 1 \\  
        0 & 0 & 1 \\  
        0 & 0 & 1 \\  
      };

      \node[lab,anchor=east,xshift=-6pt] at (M-1-1.west) {$d_1$};
      \node[lab,anchor=east,xshift=-6pt] at (M-2-1.west) {$d_2$};
      \node[lab,anchor=east,xshift=-6pt] at (M-3-1.west) {$d_3$};
      \node[lab,anchor=east,xshift=-6pt] at (M-4-1.west) {$d_4$};
      \node[lab,anchor=east,xshift=-6pt] at (M-5-1.west) {$d_5$};

      \node[lab,anchor=south] at (M-1-1.north) {$f_1$};
      \node[lab,anchor=south] at (M-1-2.north) {$f_2$};
      \node[lab,anchor=south] at (M-1-3.north) {$f_3$};
    \end{scope}

  \end{tikzpicture}
  \caption{Example Z-detector hypergraph and incidence matrix $H_Z$: detectors $d_1,\dots,d_5$ are vertices, and each fault $f_\ell$ is a (hyper)edge with support $\partial(f_\ell)$.\vspace{-10pt}}
  \label{fig:hypergraph-incidence}
\end{figure}
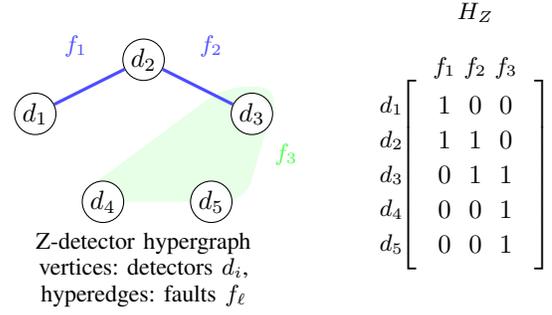

If every fault flips at most two detectors, $(\mathcal{D},\mathcal{E}_Z)$ is an
ordinary graph rather than a hypergraph. This includes many biased-noise
models in topological codes, such as XZZX and domain wall color
codes in the infinite-bias limit, and certain detector choices for
X$^3$Z$^3$. Fig.~\ref{fig:hypergraph-incidence}.

\subsection{Static strip-like biased codes}

We record two static examples whose Z sectors already exhibit the strip-wise
structure we formalise later. In the infinite-bias (pure dephasing) limit,
both the XZZX surface code~\cite{bonilla2021xzzx,bonilla2021xzzxsupp} and
the domain wall color code~\cite{tiurev2024domainwall,domainwall_supplement}
admit a partition of Z detectors into strips 
  $\mathcal{D} = \bigsqcup_{a=1}^L \mathcal{D}_a,$ 
(diagonals in XZZX, domain lines in the domain wall color code) such that

\begin{enumerate}
  \item each elementary Z fault flips at most two detectors, and these lie in a
        single strip $\mathcal{D}_a$; and
  \item on each strip, there is a stabilizer product enforcing
        \(
          \sum_{d\in\mathcal{D}_a} s_d \equiv 0 \pmod 2,
        \)
        i.e.\ an even-parity constraint on Z excitations.
\end{enumerate}

Equivalently, after permuting rows and columns the Z-detector incidence matrix
has the block form
\[
  H_Z \;\cong\; \bigoplus_{a=1}^L H^{\mathrm{rep}}_{m_a},
\]
where $H^{\mathrm{rep}}_{m_a}$ is the incidence matrix of a length-$m_a$
one-dimensional repetition chain along strip $a$. Maximum-likelihood Z
decoding therefore factorises into independent repetition decoders on each
strip, explaining the highly biased thresholds observed for these codes.
Our notion of strip-symmetric biased codes abstracts exactly this structure
and extends it to dynamical (Floquet) codes such as X$^3$Z$^3$.
\section{Strip-Symmetric Biased Codes}
\label{sec:strip}


\subsection{Strip partitions and strip symmetry}

Let $\mathcal{Q}$ be the set of physical qubits in the window and $\mathcal{D}$ a set of detectors.

\begin{definition}[Strip partition]
A strip partition consists of:
\begin{itemize}
  \item a partition of qubits $\mathcal{Q} = S_1 \sqcup \dots \sqcup S_m$ into $m$ disjoint strips, and
  \item a partition of detectors $\mathcal{D} = D_1 \sqcup \dots \sqcup D_m$ such that each detector $d\in D_j$ is supported on qubits in $S_j$.
\end{itemize}
\end{definition}

\begin{definition}[Strip-local Z faults]
Given a strip partition, a family of elementary Z faults is strip-local if, for each fault $f$, there exists a unique strip index $j$ such that $\partial(f)\subseteq D_j$.
\end{definition}

\begin{definition}[Strip-symmetric biased code]\label{def:strip-symmetric}
A static or dynamical code is \emph{strip-symmetric biased} (under pure Z noise) if it admits a strip partition $\{(S_j,D_j)\}_{j=1}^m$ such that:
\begin{enumerate}
  \item (Strip-local faults) The elementary Z faults are strip-local.
  \item (Per-strip parity) For each strip $j$ there exists a subset $\mathcal{P}_j\subseteq D_j$ such that
  \begin{equation}
    Q_j := \prod_{d\in \mathcal{P}_j} d
  \end{equation}
  is a stabilizer (or product of stabilizers) whose noiseless outcome is always $+1$.
\end{enumerate}
\end{definition}

The operators $Q_j$ express a $\ZZ_2$ 1-form symmetry along strip $j$: the product of the corresponding detector outcomes must be $+1$ in the absence of noise. Z faults must respect this symmetry, so defects on each strip have even parity.

\subsection{Block-diagonal incidence matrix and strip hypergraph}

Strip-locality immediately constrains the structure of $H_Z$.

\begin{lemma}[Block-diagonal $H_Z$]
\label{lem:block-diag}
Let a code be strip-symmetric with strips $\{(S_j,D_j)\}_{j=1}^m$. Order detectors so that the rows of $H_Z$ are grouped as $(D_1,\dots,D_m)$, and group faults by strip. Then $H_Z$ is permuted into block-diagonal form

\begin{equation}
  H_Z \sim
  \begin{bmatrix}
    H_1 & 0   & \cdots & 0 \\
    0   & H_2 & \ddots & \vdots \\
    \vdots & \ddots & \ddots & 0 \\
    0   & \cdots & 0 & H_m
  \end{bmatrix},
\end{equation}
where $H_j$ is the restriction of $H_Z$ to detectors in $D_j$ and faults whose flipped-detector sets lie in $D_j$.
\end{lemma}
\begin{corollary}[Strip-wise hypergraph decomposition]
\label{cor:hypergraph-decomp}
Under the assumptions of Lemma~\ref{lem:block-diag}, the Z-detector hypergraph decomposes as
\begin{equation}
  (\mathcal{D},\mathcal{E}_Z) = (\mathcal{D}_1,\mathcal{E}_1) \sqcup \dots \sqcup (\mathcal{D}_m,\mathcal{E}_m),
\end{equation}
where $\mathcal{D}_j=D_j$ and $\mathcal{E}_j=\{e_f: \partial(f)\subseteq D_j\}$.
\end{corollary}

Thus, the Z-decoding graph or hypergraph breaks into independent strip components.

\subsection{Factorised ML decoding and complexity}

Let $e=(e_1,\dots,e_m)$ be the binary vector of faults, grouping faults by strip, and let $s=(s_1,\dots,s_m)$ be the detector outcomes grouped similarly. We assume independent Z faults across spacetime, so $\Prob(e) = \prod_{j=1}^m \Prob(e_j)$.

\begin{theorem}[Factorised ML decoding under pure Z noise]
\label{thm:factorised-ML}
Consider a strip-symmetric biased code under pure Z noise with independent Z faults and perfect measurements. Let $H_Z$ be block-diagonal as in Lemma~\ref{lem:block-diag}. Then:
\begin{enumerate}
  \item For any syndrome $s=(s_1,\dots,s_m)$, the conditional distribution of faults factorises:
  \begin{equation}
    \Prob(e\mid s)
    \propto
    \prod_{j=1}^m
      \Prob(e_j)\,\1[H_j e_j = s_j].
  \end{equation}
  \item ML decoding factorises:
  \begin{equation}
    \hat e_{\mathrm{ML}}(s)
    =
    \bigl(\hat e^{(1)}_{\mathrm{ML}}(s_1),\dots,\hat e^{(m)}_{\mathrm{ML}}(s_m)\bigr),
  \end{equation}
  where
  \begin{equation}
    \hat e^{(j)}_{\mathrm{ML}}(s_j)\in
      \arg\max_{e_j: H_j e_j = s_j} \Prob(e_j).
  \end{equation}
\end{enumerate}
\end{theorem}


In particular, for any decoder whose runtime scales superlinearly with graph size, strip-wise decoding can offer significant speedups.

\begin{proposition}[Complexity gain for superlinear decoders]
\label{prop:complexity}
Suppose a Z decoder has runtime $T(n)=\Theta(n^\alpha)$ on a graph with $n$ detectors, for some $\alpha>1$. Let $n_j=|D_j|$ be the number of detectors on strip $j$ and $N=\sum_j n_j$ the total number of detectors. Then monolithic decoding has runtime $T_{\mathrm{mono}}(N)=\Theta(N^\alpha)$, while strip-wise decoding has runtime
\begin{equation}
  T_{\mathrm{strip}}(N)
  = \Theta\Bigl(\sum_{j=1}^m n_j^\alpha\Bigr).
\end{equation}
If all strips have comparable size, $n_j\approx N/m$, then
\begin{equation}
  T_{\mathrm{strip}}(N) \approx \frac{T_{\mathrm{mono}}(N)}{m^{\alpha-1}},
\end{equation}
i.e.\ a speedup by a factor $\Theta(m^{\alpha-1})$.
\end{proposition}

For X$^3$Z$^3$, this theorem recovers the domain-wise matching strategy for A-type detectors under pure Z noise~\cite{setiawan2025x3z3}; for XZZX and domain wall color codes, it recovers the known line-wise repetition decoding at infinite bias~\cite{bonilla2021xzzx,tiurev2024domainwall}.

\begin{figure*}[t]
  \centering
  \begin{tikzpicture}[
    scale=0.7,
    qubit/.style={circle,draw,inner sep=1.0pt,fill=white},
    stripfill/.style={fill opacity=0.15,draw opacity=0},
    lab/.style={font=\small}
  ]

    \begin{scope}[shift={(0,0)}]
    
    \fill[stripfill,blue!40,rotate around={45:(1.5,1.5)}]
    (1.5-3,1.5-0.5) rectangle (1.5+3,1.5+0.5);
    
    \fill[stripfill,blue!20,rotate around={45:(1,2)}]
    (1-3,2-0.5) rectangle (1+3,2+0.5);
    
    \foreach \x in {0,...,3} {
    \foreach \y in {0,...,3} {
      \node[qubit] at (\x,\y) {};
    }
    }
    
    \draw[blue!70,thick] (-0.5,-0.5) -- (3.5,3.5);
    \draw[blue!50,thick,dashed] (-0.5,0.5) -- (2.5,3.5);
    
    \node[lab,blue!80,anchor=west] at (1.2,3.8) {diagonal strips};
    \node[lab] at (1.5,-0.8) {(a) $\DSR(L)$};
    \end{scope}

    \begin{scope}[shift={(6,0)}]
      \foreach \x in {0,...,3} {
        \fill[stripfill,red!20]
          ({\x-0.5},-0.5) rectangle ({\x+0.5},3.5);
      }

      \foreach \x in {0,...,3} {
        \foreach \y in {0,...,3} {
          \node[qubit] at (\x,\y) {};
        }
      }

      \foreach \k in {-0.5,0.5,1.5,2.5,3.5} {
        \draw[gray!60] (\k,-0.5) -- (\k,3.5);
      }

      \node[lab,red!80,anchor=west] at (0.2,3.8) {column strips};

      \node[lab] at (1.5,-0.8) {(b) $\CSR(L)$};
    \end{scope}

    \begin{scope}[shift={(12,0)}]
      \foreach \x in {1,3} {
        \fill[stripfill,gray!20]
          ({\x-0.5},-0.5) rectangle ({\x+0.5},3.5);
      }

      \foreach \x in {0,2} {
        \fill[stripfill,green!60]
          ({\x-0.5},-0.5) rectangle ({\x+0.5},3.5);
      }

      \foreach \x in {0,...,3} {
        \foreach \y in {0,...,3} {
          \node[qubit] at (\x,\y) {};
        }
      }

      \foreach \k in {-0.5,0.5,1.5,2.5,3.5} {
        \draw[gray!60] (\k,-0.5) -- (\k,3.5);
      }

      \node[lab,green] at (0.0,3.8) {active};
      \node[lab,green] at (2.0,3.8) {active};

      \node[lab] at (1.5,-0.8) {(c) $\HCSR(L)$};
    \end{scope}

  \end{tikzpicture}
  \caption{Strip structure in synthetic benchmarks.
(a) $\DSR(L)$: diagonal strips with nearest-neighbour ZZ repetition chains.
(b) $\CSR(L)$: columns are vertical ZZ repetition strips.
(c) $\HCSR(L)$: every second column active, mimicking unshaded X$^3$Z$^3$ domains and halving detector density at fixed strip length.}
  \label{fig:dsr-csr-hcsr}
\end{figure*}
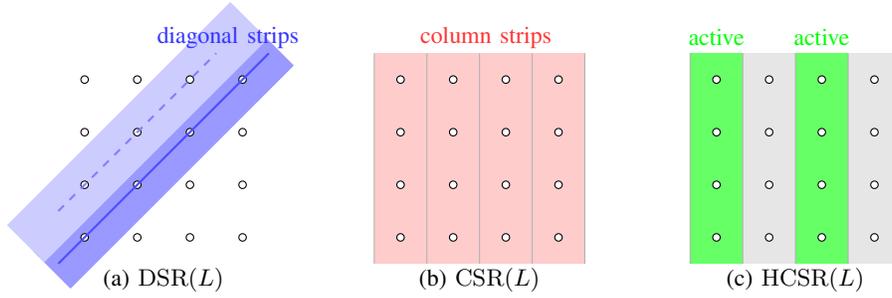

\section{
characterisation and Floquet construction}
\label{sec:design}

We recast strip symmetry as a per-strip $\ZZ_2$ 1-form symmetry and give a
general Floquet construction via domain-wise Clifford deformations.

\subsection{Per-strip $\ZZ_2$ 1-form symmetries}

Let $\mathcal{D}$ be the set of detectors and
\[
  \mathcal{D} \;=\; \bigsqcup_{j} D_j
\]
a strip partition.

\begin{definition}[Per-strip $\ZZ_2$ 1-form symmetry]
A per-strip $\ZZ_2$ 1-form symmetry is a family of Pauli observables
\[
  Q_j \;=\; \prod_{d\in\mathcal{P}_j} d, \qquad
  \mathcal{P}_j \subseteq D_j,
\]
such that each $Q_j$ is (a product of) stabilizers with noiseless outcome $+1$.
Equivalently, $Q_j$ imposes an even-parity constraint on the flipped detectors
in $D_j$.
\end{definition}

We next characterise strip symmetry algebraically.

\begin{theorem}[Strip symmetry and 1-form symmetries]
\label{thm:strip-1form}
Consider pure Z noise with a fixed set of elementary Z faults
$f\in\mathcal{F}$. For a code with detector set $\mathcal{D}$ and strip
partition $\{D_j\}$, the following are equivalent:
\begin{enumerate}
  \item The code is strip-symmetric biased (Def.~\ref{def:strip-symmetric}):
        each fault $f$ flips detectors contained in a single strip $D_j$ and
        the Z-detector incidence matrix $H_Z$ is block-diagonal after a
        strip-respecting permutation.
  \item There exists a per-strip $\ZZ_2$ 1-form symmetry $\{Q_j\}$ and every
        elementary Z fault flips either $0$ or $2$ detectors, which, when
        non-zero, lie in a common strip $D_j$.
\end{enumerate}
\end{theorem}

Thus, strip symmetry is combination of (i) fault locality in the
strip partition and (ii) a per-strip 1-form parity constraint.

\subsection{Static and Floquet examples}

We collect the motivating examples in a single statement.

\begin{proposition}[Strip symmetry of XZZX, DWCC, and X$^3$Z$^3$]
\label{prop:examples-strip}
In the infinite-bias pure Z model, the following codes are strip-symmetric
biased in the sense of Theorem~\ref{thm:strip-1form}:
\begin{enumerate}[(a)]
  \item \emph{XZZX surface code}: plaquette detectors, strips given by lattice
        diagonals~\cite{bonilla2021xzzx,bonilla2021xzzxsupp};
  \item \emph{Domain wall color code}: face detectors, strips given by
        domain-wall-oriented lines~\cite{tiurev2024domainwall,domainwall_supplement};
  \item \emph{X$^3$Z$^3$ Floquet code}: A-type plaquette detectors on
        unshaded vertical domains, strips given by these
        domains~\cite{setiawan2025x3z3,setiawan2025x3z3supp}.
\end{enumerate}
In each case, every single-qubit Z error flips two detectors on a single
strip, and products of stabilizers (or detector operators) along each strip
give per-strip 1-form symmetries $Q_j$ with noiseless outcome $+1$.
\end{proposition}

\subsection{Strip-symmetric Floquet codes via domain-wise Cliffords}

We now describe a general way to generate strip-symmetric Floquet codes
starting from a CSS Floquet code and a strip partition of its qubits.

Let $C$ be a dynamical code with qubit set $\mathcal{Q}$, measurement sets
$\{M_i\}$, and a strip partition
\[
  \mathcal{Q} \;=\; \bigsqcup_{j=1}^m S_j.
\]

\begin{definition}[Domain-wise Clifford deformation]
\label{def:domain-wise-clifford}
Choose single-qubit Clifford unitaries $U_j$ (up to phase) and define
\[
  U \;=\; \bigotimes_{j=1}^m U_j^{\otimes |S_j|},
\]
i.e.\ $U$ applies $U_j$ to every qubit in strip $S_j$.
The domain-wise Clifford-deformed code $C' = U C U^\dagger$ has measurement
sets
\[
  M'_i \;=\; \{ U M U^\dagger : M \in M_i \}
\]
and inherits the same strip partition $\{S_j\}$. The deformation is
\emph{bias-shifting} if each $U_j \in \{\mathbb{I},H,HS\}$ up to phase.
\end{definition}
\begin{proposition}[Distance under Clifford conjugation]
\label{prop:distance-clifford}
Let $C$ be a dynamical code in the sense of Fu and Gottesman~\cite{fu2025dynamical}
with unmasked distance $d_{\mathrm{unmask}}(C)$, and let $C' = U C U^\dagger$
for a Clifford $U$. Then
\[
  d_{\mathrm{unmask}}(C') \;=\; d_{\mathrm{unmask}}(C).
\]
\end{proposition}

This holds because Clifford conjugation induces a symplectic automorphism of
the Pauli group that preserves commutation and weight.

We can now combine strip symmetry in a convenient Pauli basis with a
bias-shifting domain-wise Clifford.

\begin{theorem}[Strip-symmetric Floquet deformation]
\label{thm:strip-floquet-deformation}
Let $C$ be a CSS Floquet code on a 3-colourable trivalent lattice with
measurement schedule $\{M_i\}$ and strip partition $\{S_j\}$. Assume that, in
some Pauli basis, the Z-detector structure of $C$ is strip-symmetric biased as
in Theorem~\ref{thm:strip-1form}: there exist per-strip 1-form symmetries
$Q_j$ supported on $S_j$, and each single-qubit Z error flips at most two
detectors in a single strip. Let $U$ be a bias-shifting domain-wise Clifford
and define $C' = U C U^\dagger$. Then, under the physical pure Z noise model,
$C'$ is a strip-symmetric biased dynamical code, and maximum-likelihood Z
decoding for $C'$ factorises exactly across strips as in
Theorem~\ref{thm:factorised-ML}.
\end{theorem}

\begin{remark}[X$^3$Z$^3$ as a special case]
In X$^3$Z$^3$~\cite{setiawan2025x3z3}, $C$ is the honeycomb CSS Floquet code,
$\{S_j\}$ are the unshaded vertical domains, and $U_j\in\{\mathbb{I},H\}$
depending on the domain. Proposition~\ref{prop:examples-strip} and
Theorem~\ref{thm:strip-floquet-deformation} then show that the resulting
Floquet code is strip-symmetric biased at infinite Z bias with respect to
A-type detectors, justifying the domain-wise MWPM decoding strategy of
\cite{setiawan2025x3z3,setiawan2025x3z3supp}.
\end{remark}

\section{Strip Benchmarks and Numerics}
\label{sec:benchmarks}
We instantiate Def.~\ref{def:strip-symmetric} and Thm.~\ref{thm:strip-1form} on three physical families (XZZX, domain wall color code (DWCC), X$^3$Z$^3$) and three synthetic strip benchmarks (diagonal-, column-, and half-density column-strip repetition, $\DSR$, $\CSR$, $\HCSR$). In each case we partition qubits into diagonals (XZZX, $\DSR$), domain lines (DWCC, $\CSR$), or vertical domains (X$^3$Z$^3$, $\HCSR$)---see Fig.~\ref{fig:dsr-csr-hcsr}, and place ZZ detectors as 1D chains along each strip so that (i) every single-Z fault flips at most two detectors on one strip and (ii) the product of all detectors on a strip is a stabilizer, enforcing the per-strip 1-form parity constraint and the block-diagonal $H_Z$ of Lemma~\ref{lem:block-diag}. For XZZX, DWCC, and X$^3$Z$^3$ this recovers the strip structure of Prop.~\ref{prop:examples-strip} (and can be obtained from CSS/Floquet parents via Thm.~\ref{thm:strip-floquet-deformation}), while the synthetic models are pure Z-detector ``shadows'': exact stacks of 1D repetition chains tailored to the setting of Thm.~\ref{thm:factorised-ML} with matched detector/fault counts.

For each code and size $L$, we  
extract the Z-detector incidence matrix $H_Z$, permute rows and columns to group detectors and faults by strip, and record $m$ (strips with at least one detector), $\min_j|D_j|,\max_j|D_j|$ (min/max detectors per strip), \emph{off-block} (nonzeros of $H_Z$ outside strip-diagonal blocks), \emph{non-local} (faults hitting multiple strips), and totals $N_{\mathrm{det}},N_{\mathrm{fault}}$. Table~\ref{tab:block-stats-data} shows these statistics for $L=3,4,5$. For all six families \emph{off-block}$=0$ and \emph{non-local}$=0$, so $H_Z$ is block-diagonal and all Z faults are strip-local, and the synthetic benchmarks reproduce the $(m,|D_j|,N_{\mathrm{det}},N_{\mathrm{fault}})$ patterns of their matched physical codes (XZZX$\leftrightarrow\DSR$, DWCC$\leftrightarrow\CSR$, X$^3$Z$^3\leftrightarrow\HCSR$), confirming that they are idealised strip-symmetric Z-detector shadows.

\begin{table}[t]
  \centering
  \caption{Strip statistics for idealised Z-detector models at $L=3,4,5$.}
  \setlength{\tabcolsep}{2pt}
  \label{tab:block-stats-data}
  \begin{tabular}{lrrrrrrrr}
    \toprule
    Code & $L$ & $m$ & $\min_j |D_j|$ & $\max_j |D_j|$ &
      off-block & non-local & $N_{\mathrm{det}}$ & $N_{\mathrm{fault}}$ \\
    \midrule
XZZX & 3 & 3 & 1 & 2 & 0 & 0 & 4 & 9 \\
XZZX & 4 & 5 & 1 & 3 & 0 & 0 & 9 & 16 \\
XZZX & 5 & 7 & 1 & 4 & 0 & 0 & 16 & 25 \\
DWCC & 3 & 3 & 2 & 2 & 0 & 0 & 6 & 9 \\
DWCC & 4 & 4 & 3 & 3 & 0 & 0 & 12 & 16 \\
DWCC & 5 & 5 & 4 & 4 & 0 & 0 & 20 & 25 \\
X3Z3 & 3 & 2 & 2 & 2 & 0 & 0 & 4 & 9 \\
X3Z3 & 4 & 2 & 3 & 3 & 0 & 0 & 6 & 16 \\
X3Z3 & 5 & 3 & 4 & 4 & 0 & 0 & 12 & 25 \\
DSR & 3 & 3 & 1 & 2 & 0 & 0 & 4 & 9 \\
DSR & 4 & 5 & 1 & 3 & 0 & 0 & 9 & 16 \\
DSR & 5 & 7 & 1 & 4 & 0 & 0 & 16 & 25 \\
CSR & 3 & 3 & 2 & 2 & 0 & 0 & 6 & 9 \\
CSR & 4 & 4 & 3 & 3 & 0 & 0 & 12 & 16 \\
CSR & 5 & 5 & 4 & 4 & 0 & 0 & 20 & 25 \\
HCSR & 3 & 2 & 2 & 2 & 0 & 0 & 4 & 9 \\
HCSR & 4 & 2 & 3 & 3 & 0 & 0 & 6 & 16 \\
HCSR & 5 & 3 & 4 & 4 & 0 & 0 & 12 & 25 \\
    \bottomrule
  \end{tabular}
\end{table}

\begin{figure}[t]
  \centering
  \includegraphics[width=\columnwidth]{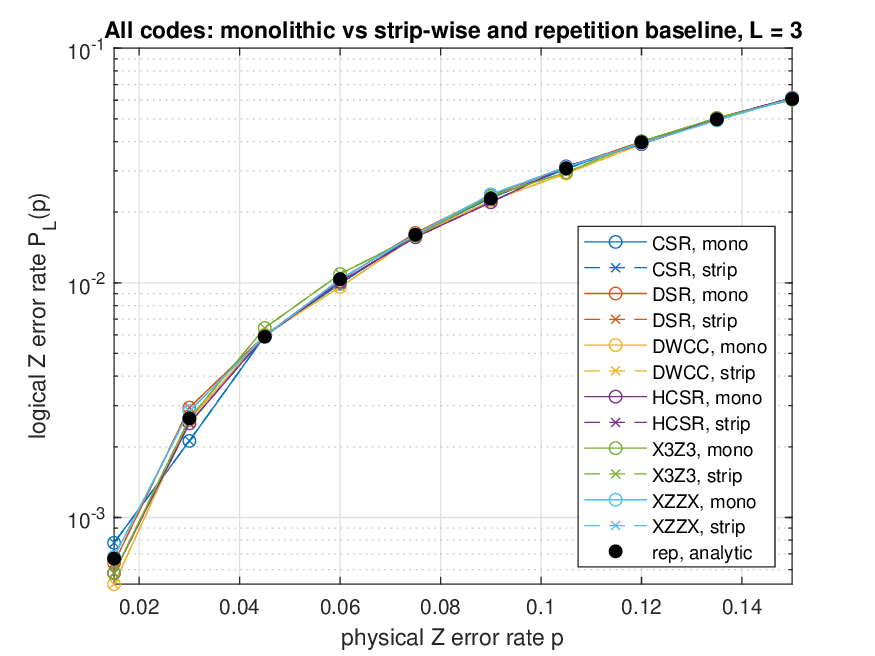}
  \caption{Logical Z error rate \(P_L(p)\) vs.\ physical Z error rate \(p\).}
  \label{fig:PL-rep-threshold}
\end{figure}

On any strip that supports a logical Z, the Z-sector reduces to a length-$L$
classical repetition chain under independent bit flips of probability $p$,
with ML logical error rate
\begin{equation}
  P_L^{\mathrm{rep}}(p)
    = \sum_{w=\lceil(L+1)/2\rceil}^{L}
        \binom{L}{w} p^w (1-p)^{L-w}.
  \label{eq:rep-PL}
\end{equation}
Fig.~\ref{fig:PL-rep-threshold} shows $P_L(p)$ for XZZX, DWCC, X$^3$Z$^3$
and their synthetic shadows under pure Z code-capacity noise, using an exact
ML Z decoder (minimum-weight bit-flip decoding via exhaustive search) both
monolithically and strip-wise. For all cases the numerical points lie on the
analytic repetition baseline $P_L^{\mathrm{rep}}(p)$ within Monte Carlo error
and the monolithic and strip-wise curves coincide, confirming
Theorem~\ref{thm:factorised-ML} and a code-capacity Z-threshold consistent
with $p_{\mathrm{th}}=1/2$.

\section{Summary and Design Outlook}
\label{sec:conclusion}
We introduced \emph{strip-symmetric biased codes}, a structural framework
capturing the strip-like Z-syndrome organisation that underlies the
biased-noise performance of XZZX, domain wall color codes, and X$^3$Z$^3$:
strip-local Z faults and per-strip stabilizer products (a $\mathbb{Z}_2$
1-form symmetry) make the Z-detector incidence matrix block diagonal so ML Z
decoding factorises across strips with matching complexity gains. Compared to
code-specific constructions such as XZZX, the domain wall color code, and
X$^3$Z$^3$, our strip-symmetric framework isolates this mechanism at the level
of detector hypergraphs, independently of any particular lattice or schedule.
This has two main advantages: (i) general ML-factorisation and
decoding-complexity guarantees that apply to \emph{any} code realising strip
symmetry, rather than being proved case by case; and (ii) it turns
line/domain symmetries from an observed property of a few hand-crafted codes
into a design primitive, allowing one to choose strip partitions, per-strip
codes, detector densities, and domain-wise Clifford deformations to generate
and benchmark whole families of static and Floquet codes with inherited
biased-decoding benefits. Our strip-symmetric Floquet deformation theorem and
synthetic detector families $\DSR(L)$, $\CSR(L)$, and $\HCSR(L)$ illustrate
this design perspective and provide analytically tractable benchmarks and
idealised Z-detector shadows for future 
constructions.

\newpage
\bibliographystyle{IEEEtran}
\bibliography{refs}

\newpage
\appendix
\section*{Proof of Lemma \ref{lem:block-diag}}
\begin{proof}
By strip-locality, each fault $f$ has $\partial(f)\subseteq D_j$ for a unique $j$. If we order rows by strips and group columns accordingly, each column has nonzero entries only within the row group corresponding to its strip. This yields the block-diagonal form.
\end{proof}

\section*{Proof of Corollary \ref{cor:hypergraph-decomp}}
\begin{proof}
Each hyperedge $e_f$ is contained in exactly one $D_j$, so the hypergraph is a disjoint union of these components.
\end{proof}

\section*{Proof of Theorem \ref{thm:factorised-ML}}
\begin{proof}
By Bayes' rule,
\begin{equation}
  \Prob(e\mid s)\propto \Prob(e)\,\1[H_Z e = s].
\end{equation}
Independence of faults across strips yields $\Prob(e) = \prod_j \Prob(e_j)$. Block diagonality implies $H_Z e = s$ iff $H_j e_j = s_j$ for all $j$. Thus
\begin{equation}
  \Prob(e\mid s)
  \propto \prod_{j=1}^m
            \Prob(e_j)\,\1[H_j e_j = s_j],
\end{equation}
proving the first claim. 

To maximise the product over $e$ subject to the constraints, we can maximise each factor independently over $e_j$ satisfying $H_j e_j = s_j$, which yields the factorised ML solution.
\end{proof}

\section*{Proof of Proposition \ref{prop:complexity}}
\begin{proof}
The strip-wise runtime is the sum of per-strip runtimes:
\(
  T_{\text{strip}}(N) = \sum_j \Theta(n_j^\alpha).
\)
The monolithic runtime is $T_{\text{mono}}(N)=\Theta(N^\alpha)$. For balanced strips,
\(
  \sum_j n_j^\alpha \approx m (N/m)^\alpha = N^\alpha/m^{\alpha-1}.
\)
\end{proof}

\section*{Proof of Theorem \ref{thm:strip-1form}}
\begin{proof}
We work over $\F_2$.  Let $\mathcal{D}$ be the detector set and
$\mathcal{F}$ the fixed set of elementary $Z$ faults.  Define the incidence
matrix $H_Z\in \F_2^{\mathcal{D}\times \mathcal{F}}$ by
\[
  (H_Z)_{d,f}=1 \quad\Longleftrightarrow\quad \text{fault $f$ flips detector $d$},
\]
and let $h_f\in\F_2^{\mathcal{D}}$ denote the $f$th column.  A fault pattern
$e\in \F_2^{\mathcal{F}}$ produces a detector-flip (syndrome) pattern
$s=\partial_Z(e):=H_Z e\in \F_2^{\mathcal{D}}$.

Fix a strip partition $\mathcal{D}=\bigsqcup_{j=1}^m D_j$.  Let
$q_j\in\F_2^{\mathcal{D}}$ be the indicator of $D_j$, i.e.
$(q_j)_d=1$ iff $d\in D_j$.
Define also the strip of a fault (when it exists) by the condition
$\mathrm{supp}(h_f)\subseteq D_{\sigma(f)}$.

\medskip
\noindent\emph{1-form symmetry in the incidence formalism.}
Given detectors as $\pm1$-valued observables $\{d\}_{d\in\mathcal{D}}$ with
noiseless outcomes $+1$, define for each strip
\[
  Q_j \;:=\; \prod_{d\in D_j} d .
\]
Then $Q_j$ has noiseless outcome $+1$.  Under an elementary fault $f$, the
value of $Q_j$ flips iff the number of flipped detectors in $D_j$ is odd,
i.e.\ iff $q_j^\top h_f = 1$.  Hence $Q_j$ is a conserved $\ZZ_2$ symmetry
under all elementary faults iff
\begin{equation}
  q_j^\top H_Z \;=\; 0 \quad\text{in }\F_2^{\mathcal{F}}.
  \label{eq:qj-left-kernel}
\end{equation}
This is exactly the statement that every elementary fault flips an even
number of detectors in strip~$D_j$.

\medskip
\noindent\textbf{(2) $\Rightarrow$ (1).}
Assume (2).  By hypothesis, for each fault $f$ either $h_f=0$ or
$|\mathrm{supp}(h_f)|=2$ and $\mathrm{supp}(h_f)\subseteq D_j$ for some $j$.
Define $\sigma(f)=j$ when $h_f\neq 0$ and $\sigma(f)=0$ when $h_f=0$.  Now
permute the detector rows so that detectors are grouped by strips
$(D_1,\dots,D_m)$, and permute the fault columns so that faults are grouped
by their label $\sigma(f)\in\{0,1,\dots,m\}$.  Because each nonzero column
has support entirely within a single strip, no column has a $1$ entry outside
its strip block.  Therefore the permuted matrix is block diagonal (with an
additional all-zero block for $\sigma(f)=0$ columns if present), i.e.\
$H_Z$ is block-diagonal after a strip-respecting permutation.  This is
precisely condition~(1).

\medskip
\noindent\textbf{(1) $\Rightarrow$ (2).}
Assume (1).  By the strip-locality part, for each fault $f$ there exists a
strip index $\sigma(f)$ such that $\mathrm{supp}(h_f)\subseteq D_{\sigma(f)}$
(or $h_f=0$).  By the block-diagonal part, after strip-respecting
permutations $P_D,P_F$ we may write
\[
  P_D H_Z P_F \;=\; \mathrm{diag}(H_1,\dots,H_m),
\]
where $H_j$ is the restriction to rows $D_j$ and columns $F_j$ (faults
supported on strip $j$).  Consider the operator
$Q_j=\prod_{d\in D_j} d$.  Its noiseless outcome is $+1$.
Moreover, since $f$ is supported on a \emph{single} strip, it can only affect
$Q_{\sigma(f)}$; for all $k\neq \sigma(f)$ we have $q_k^\top h_f=0$.

In the pure-$Z$ incidence setting of this theorem, elementary faults are
assumed to be \emph{pair-creating} on detectors, i.e.\ each $h_f$ has Hamming
weight $0$ or $2$.\footnote{Equivalently, $H_Z$ is the incidence matrix of a
graph (not a general hypergraph), as in the code-capacity pure-$Z$ models for
XZZX, DWCC, and X$^3$Z$^3$ discussed in
Proposition~\ref{prop:examples-strip}.  If elementary faults can flip larger
even numbers of detectors (hyperedges), the statement adapts by replacing
``$0$ or $2$'' with ``even'', and the proof below remains identical.}
Hence, whenever $h_f\neq 0$ we have $|\mathrm{supp}(h_f)|=2$ and
$\mathrm{supp}(h_f)\subseteq D_{\sigma(f)}$.  In particular, for each strip
$j$ and each fault $f$,
\[
  q_j^\top h_f \;=\; |\mathrm{supp}(h_f)\cap D_j|\;\;(\mathrm{mod}\;2)
  \;\in\;\{0,2\}\equiv 0\pmod 2,
\]
so \eqref{eq:qj-left-kernel} holds for all $j$.  Therefore each $Q_j$ commutes
with every elementary fault and has noiseless outcome $+1$, i.e.\ $\{Q_j\}$
is a per-strip $\ZZ_2$ 1-form symmetry.  This also establishes that every
elementary fault flips either $0$ or $2$ detectors, and when nonzero those
detectors lie in a common strip $D_{\sigma(f)}$.  Thus (2) holds.

\medskip
This proves the equivalence of (1) and (2).
\end{proof}

\section*{Proof of Proposition \ref{prop:examples-strip}}
\begin{proof}
Let $D$ denote the set of Z-type \emph{detectors} being measured (plaquette or
face detectors in the static cases; A-type plaquette detectors in the Floquet
case), and let $F$ denote the set of \emph{elementary Z faults} (single-qubit
$Z$ errors, including a time index if the code is dynamical).  For each
$f\in F$ define its detector-incidence vector
\[
  h(f)\in \mathbb{F}_2^{D},\qquad
  h(f)_d := 
  \begin{cases}
    1, & \text{if } \{f,d\}=0 \ (\text{$f$ flips detector $d$}),\\
    0, & \text{otherwise}.
  \end{cases}
\]
Stacking the vectors $\{h(f)\}_{f\in F}$ gives the Z-detector incidence matrix
$H_Z\in \mathbb{F}_2^{D\times F}$ (columns indexed by faults, rows by
detectors).  Under pure Z code-capacity noise and perfect measurements, the
(binary) detector flip pattern caused by an error $e\in\mathbb{F}_2^{F}$ is
$s = H_Z e$.

\medskip
\noindent\textbf{Abstract strip-symmetry criterion.}
Suppose there exists a strip labelling $\pi:D\to[m]$ such that for each fault
$f\in F$ there is some $j\in[m]$ with
\begin{equation}
  \supp h(f)\subseteq D_j := \pi^{-1}(j)
  \qquad\text{and}\qquad
  |\supp h(f)| \in \{0,2\}.
  \label{eq:strip-local-2}
\end{equation}
(If a fault flips a single physical detector on a boundary, one may include a
standard \emph{virtual boundary detector} $d_{\partial,j}$ with fixed outcome
$+1$ in the same strip so that~\eqref{eq:strip-local-2} holds with
$|\supp h(f)|=2$ in the augmented detector set.)
Then, after permuting rows by $\pi$ (and permuting columns by any compatible
fault-to-strip labelling), $H_Z$ is block-diagonal with blocks indexed by
$j\in[m]$, i.e.\ strip-locality.

Moreover, define the strip-parity vectors $u_j\in\mathbb{F}_2^{D}$ by
$(u_j)_d=1$ if $d\in D_j$ and $0$ otherwise.  By~\eqref{eq:strip-local-2},
for any fault $f$ we have
\[
  u_j^\top h(f) \;=\; |\supp h(f)\cap D_j|\;\equiv\;0 \pmod 2,
\]
hence
\begin{equation}
  u_j^\top H_Z \;=\; 0 \quad\text{in }\mathbb{F}_2^{F}.
  \label{eq:left-kernel}
\end{equation}
Thus each strip parity $u_j^\top s$ is \emph{conserved} under Z faults.  In
operator language, letting
\[
  Q_j \;:=\; \prod_{d\in D_j} d ,
\]
we have for any elementary fault $f$ that
\[
  f Q_j \;=\; (-1)^{\,u_j^\top h(f)} Q_j f \;=\; Q_j f,
\]
so $Q_j$ commutes with all elementary Z faults; and since $Q_j$ is a product
of measured detector operators, it has noiseless outcome $+1$.  Hence $\{Q_j\}$
realize per-strip $\mathbb{Z}_2$ 1-form symmetries in the sense of
Theorem~\ref{thm:strip-1form}.  It therefore suffices to verify
\eqref{eq:strip-local-2} and the existence of the corresponding strip
relations for each code family.

\medskip
\noindent\textbf{(a) XZZX surface code.}
In the diagonal basis of~\cite{bonilla2021xzzx,bonilla2021xzzxsupp}, Z errors
create/annihilate pairs of plaquette defects confined to a single family of
parallel lattice diagonals, and a product of plaquette stabilizers along each
such diagonal yields a constraint of fixed (noiseless) value $+1$
(diagonal-wise parity conservation).  Taking $D_j$ to be the set of plaquette
detectors on the $j$th diagonal gives~\eqref{eq:strip-local-2} and hence
\eqref{eq:left-kernel}.

\medskip
\noindent\textbf{(b) Domain wall color code.}
Refs.~\cite{tiurev2024domainwall,domainwall_supplement} show that at infinite
Z-bias the domain wall color code decouples into repetition-like degrees of
freedom along domain-wall-oriented lines: each elementary Z error produces
two face-syndrome excitations confined to a single such line, and the product
of face detectors along each line is constrained to $+1$ (line-wise parity
conservation).  Taking $D_j$ to be the face detectors on the $j$th
domain-wall-oriented line yields~\eqref{eq:strip-local-2} and thus
\eqref{eq:left-kernel}.

\medskip
\noindent\textbf{(c) X$^3$Z$^3$ Floquet code.}
Refs.~\cite{setiawan2025x3z3,setiawan2025x3z3supp} identify A-type plaquette
detectors supported on the unshaded vertical domains and prove a domain-wise
conservation law in the infinite-bias regime: each elementary Z fault flips
two A-type detectors within a single unshaded domain, and the product of all
A-type detectors within any given domain has fixed outcome $+1$.  Taking
$D_j$ to be the A-type detectors within the $j$th unshaded vertical domain
again yields~\eqref{eq:strip-local-2} and \eqref{eq:left-kernel}.

\medskip
In all three cases we obtain a strip partition $\{D_j\}$ for which every
single-qubit Z fault flips exactly two detectors within a single strip (up to
the standard boundary-vertex convention) and for which the strip products
$Q_j=\prod_{d\in D_j} d$ have noiseless outcome $+1$.  By the criterion above,
the corresponding $\{Q_j\}$ are per-strip $\mathbb{Z}_2$ 1-form symmetries,
and the codes are strip-symmetric biased in the sense of
Theorem~\ref{thm:strip-1form}.
\end{proof}

\section*{Proof of Theorem \ref{thm:strip-floquet-deformation}}
\begin{proof}
By assumptions~(1) and~(2), in some Pauli basis the code $C$ is strip-symmetric
biased under a pure Z model: there is a strip partition $(S_j,D_j)$ of qubits
and detectors, each single-qubit Z error flips either zero or two detectors in
a single $D_j$, and for each $j$ there exists a detector product
$Q_j = \prod_{d\in\mathcal{P}_j} d$ with $\mathcal{P}_j \subseteq D_j$ that is
a stabiliser of $C$ with noiseless outcome $+1$, enforcing even parity of
defects on strip $j$ (Theorem~\ref{thm:strip-1form}).

The domain-wise Clifford $U$ has the form
\[
  U = \bigotimes_{v\in S_1} U_1 \otimes \cdots \otimes \bigotimes_{v\in S_m} U_m,
\]
where each $U_j$ is a single-qubit Clifford. Conjugation by $U$ maps:
\begin{itemize}
  \item each detector $d\in D_j$ to $d' = U d U^\dagger$ supported on the same
        strip $S_j$, so we may take $D'_j := \{d': d\in D_j\}$ as the detectors
        for $C'$;
  \item each per-strip stabiliser product $Q_j$ to
        $Q'_j := U Q_j U^\dagger = \prod_{d\in\mathcal{P}_j} d'$, which is a
        stabiliser of $C'$ with noiseless outcome $+1$ and support contained in
        $D'_j$.
\end{itemize}

Now consider the infinite-bias pure Z noise model on $C'$: each physical Z
error $Z_v$ acts on the deformed code $C'$; in the original code $C$ this
corresponds to the Pauli
\[
  \tilde{E}_v := U^\dagger Z_v U,
\]
which is a single-qubit Pauli ($\pm X_v,\pm Y_v$ or $\pm Z_v$) because $U$ is
a tensor product of single-qubit Cliffords. By construction of the
bias-shifting deformation, we choose the effective Z axis for $C$ so that the
dominant error on each qubit is precisely $\tilde{E}_v$. In that effective
basis, assumptions~(1) and~(2) say that each $\tilde{E}_v$ flips at most two
detectors in a single $D_j$ and commutes with $Q_j$.

Conjugating back, the set of detectors flipped by $Z_v$ in $C'$ is exactly
$\{d' : d\in\partial(\tilde{E}_v)\} \subseteq D'_j$, so elementary Z faults in
$C'$ are strip-local w.r.t.\ $(S_j,D'_j)$. Moreover,
\[
  Q'_j Z_v Q_j'^\dagger
  = U Q_j U^\dagger Z_v U Q_j^\dagger U^\dagger
  = U Q_j \tilde{E}_v Q_j^\dagger U^\dagger
  = U \tilde{E}_v U^\dagger
  = Z_v,
\]
because $Q_j$ commutes with $\tilde{E}_v$. Hence $Q'_j$ commutes with all Z
errors, so the product of detector outcomes in $\mathcal{P}_j$ remains $+1$
for any Z-error pattern on $C'$, and the number of defects on each strip is
even.

Thus $C'$ satisfies the two conditions of
Definition~\ref{def:strip-symmetric}: strip-local Z faults and per-strip
stabiliser products enforcing even parity. Therefore $C'$ is strip-symmetric
biased under the physical pure Z model. By Lemma~\ref{lem:block-diag} and
Theorem~\ref{thm:factorised-ML}, the Z-detector incidence matrix and
Z-detector hypergraph for $C'$ decompose into strip-wise blocks and ML Z
decoding factorises exactly across strips.
\end{proof}

\end{document}